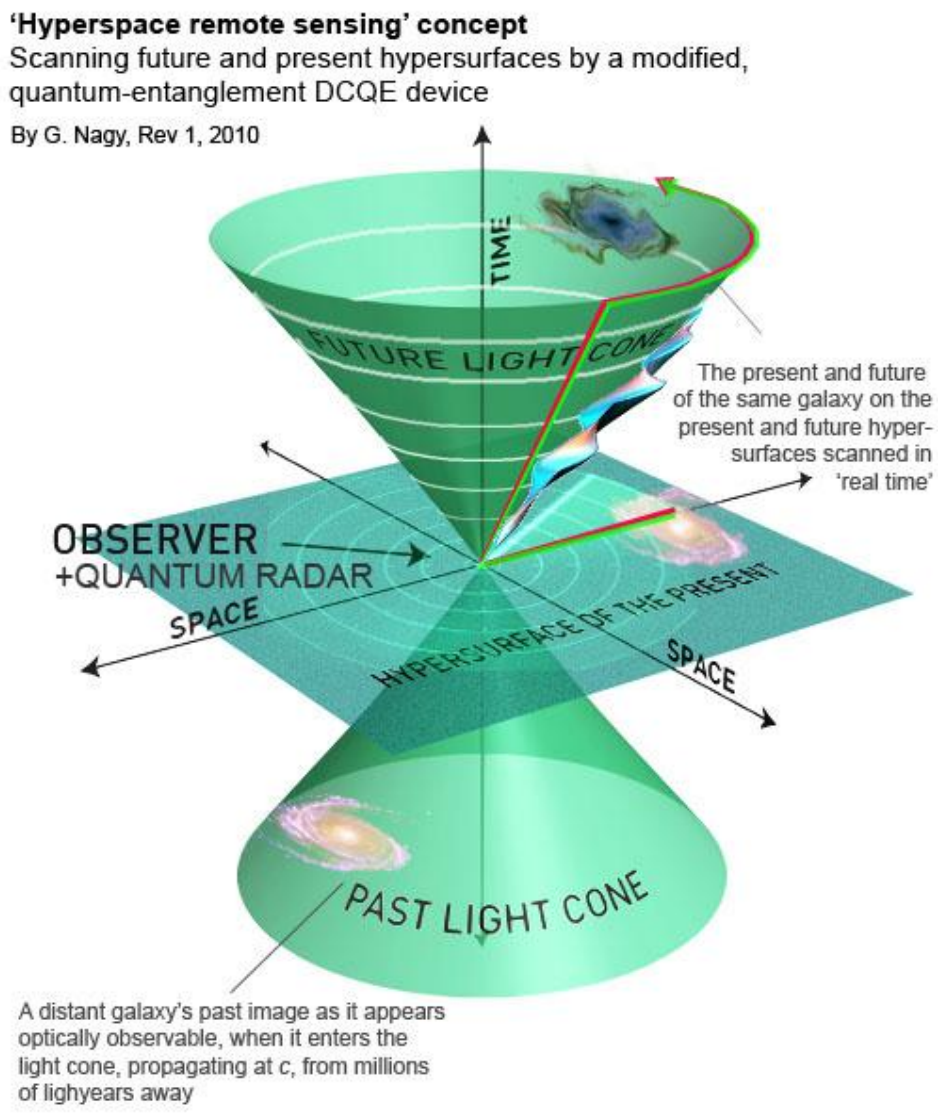

# Remote-Sensing Quantum Hyperspace by Entangled Photon Interferometry

*By Gergely A. Nagy, Rev. 1.5, 2011*

| | |
|---|---|
| **Sc.Field** | Quantum Physics / Quantum optics |
| **Topic** | Hyperspace interactions |

# Remote-Sensing Quantum Hyperspace by Entangled Photon Interferometry

By *Gergely A. Nagy*, Rev 1.5, 2011

### Abstract

Even though ideas of extracting future-related, or Faster-Than-Light (FTL) information from hyperspace using quantum entanglement have generally been refuted in the last ten years, in this paper we show that the original 'Delayed Choice Quantum Eraser Experiment', 1st performed by Yoon-Ho Kim, R. Yu, S.P. Kulik, Y.H. Shih, designed by Marlan O. Scully & Drühl in 1982-1999, still features various hidden topological properties that may have been overlooked by previous analysis, and which prohibit, by principle, such extraction of future-related or real-time information from the detection of the signal particle on the delayed choice of its entangled idler twin(s).

We show that such properties can be removed, and quantum-level information from certain hypersurfaces of past, present or future spacetime may be collected real-time, without resulting in any paradox or violation of causality. We examine the possible side effects of the 'Multi-Dimensional Hyperwaves Theory' (also presented as an appendix to this paper), on all above implementations.

### Original experiment interpretation

Yoon-Ho Kim, R. and by Marlan O. Scully[1] had shown that it is possible to delay both erasing and marking which-way path information using entangled photons by SPCD separation in any such entanglement- combined double-slit experiment(s). Delay, or distance was not limited (in time, or space either).

The experiment setup is shown in Fig. 1.

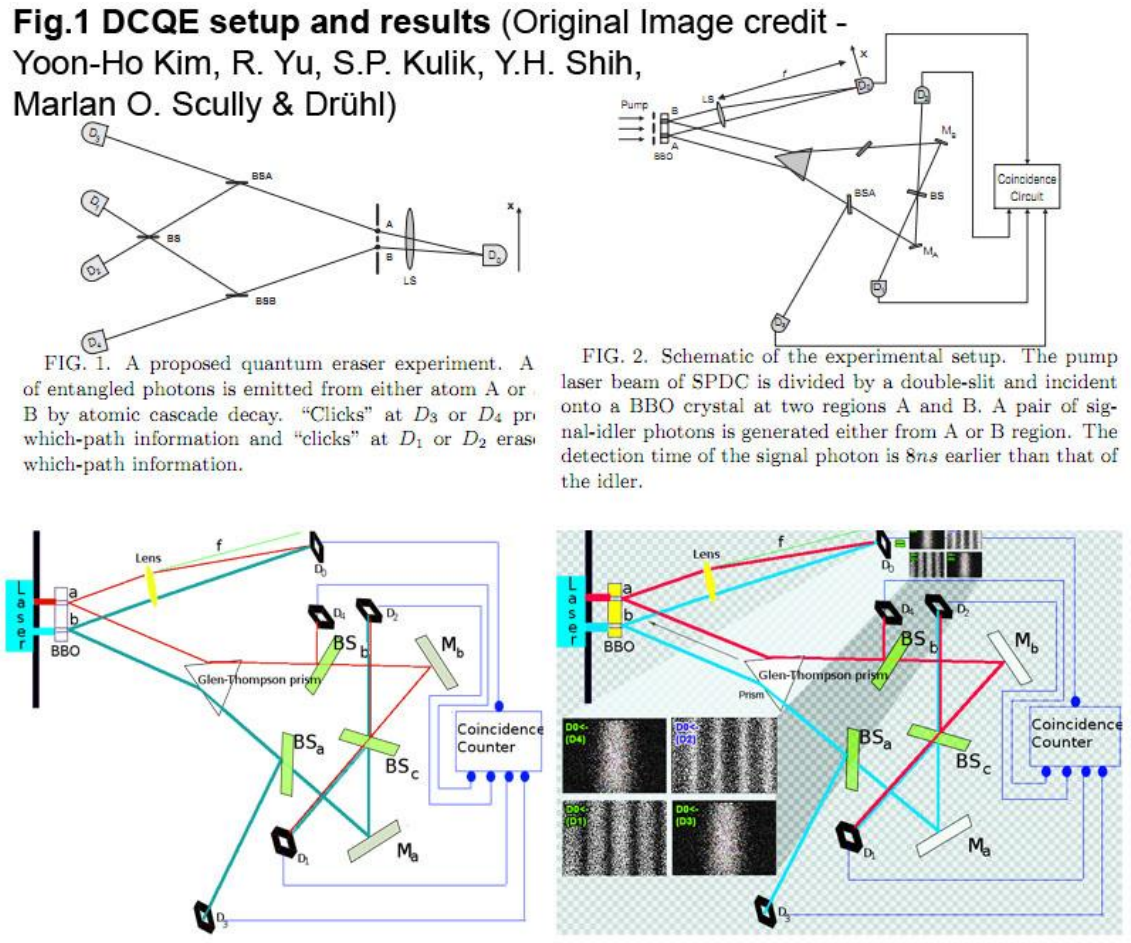

**Fig.1 DCQE setup and results** (Original Image credit - Yoon-Ho Kim, R. Yu, S.P. Kulik, Y.H. Shih, Marlan O. Scully & Drühl)

The experiment seems to have proven that the earlier detection of the (signal) photon was always – or at least statistically - correlated with the future choice of its idler twin(s).

Therefore it also appears to have proven that entanglement between the particles and their twins was not only independent of space-like separation, but also independent of time *[in the local observer's frame of reference]*.

### A relativistic interpretation

We propose an alternative theory to explain the phenomenon. If we presume that both signal and the idler photons propagate at the speed of light ($c$)[1], in their frame of reference, the theory of Relativity[2] implies that time, for them, is not needed to reach either ($D_0$-$D_n$) screens, as they are equally close to the point of SPCD (zero distance). So the difference between the length of optical paths is irrelevant (for the photon). Both the signal and the idler reach their destinations (their detection screens) in no time, having travelled through all mirrors, beamsplitters whichever they will meet in the local observer's frame of reference, *even if* the experiment setup is intentionally changed between $D_0$ 'signal' and $D_n$ 'idler' detections (which is obviously possible in the local observer's frame of reference).

So, what we may measure as 'time of detection', or delay between the detection of the signal and the idler, only exists for the local observer. Photons are reaching their target at the same time (in their frame of reference), so they indeed can, and *do* 'know' instantly how they will be detected (which-way path marked or erased)[2].

---

[1] Another possibility is that only the sum / avg of $V_s$ and $V_i$ equals $c$, and either propagations happen at speeds>$c$, (in the local observers frame of reference), as accurate detection of position of either particles may result in extreme uncertainty in its twin's speed (having no mass, or momentum), based on an extension of Heisenberg's uncertainty principle. Both alternatives are discussed in detail later in this article.

[2] Note that this implies an already existing future; but does not imply that there can be only one future. It would be very much consistent with the 'Many-Worlds Theory'[3] , that the detection at $D_0$ collapses the wavefunction only for the local observer's universe, and a new universe would be spawned in which detection in $D_0$ could be different, implicating a different future (and not resulting in determinism).

However, extracting instantly available, future-related information on the delayed choice of the signal photon's entangled twin, using data only from the signal ($D_0$) screen, for the local observer was theorized to be impossible without involving a 'coincidence counter' device, which could remotely match-and filter the entangled idler's wave-function collapses on the remote screens, ($D_1$-$D_2$, $D_3$-$D_4$ detectors, respectively). And, as that 'coincidence counter' could only be accessed by the speed of light, early access to results would be prohibited[3].

The restriction manifested itself in the $D_1$-$D_2$ detection screen's interference fringes being phase-shifted by exactly 180°, or $\pi$, thus canceling each other out to a collapsed waveform, making it undistinguishable from the $D_3$-$D_4$ detections.

Therefore, by a detection in $D_0$ it could never be pre-assumed whether it will be contributing to QM erasing (both-ways $D_1$, $D_2$) or to a marking (which-way $D_3$, $D_4$) joint detection.

With such constrains, it seemed to have been proven to be impossible, at least with the topology used, to obtain the information before future detection of the signal's idler twin as well.

## Phase shift development

The 180° ($\pi$) phase-shift in $D_1$ and $D_2$ complementary interference patterns can either be explained by QM mathematics (as Yoon-Ho Kim, R. and by Marlan O. Scully had shown in their paper), or simply by the redundant topological symmetry of the detectors in the idler part of the experiment (i.e. trying to extract the both-ways or no-path information with 2 independent detectors, mirroring them symmetrically, leaving the chance for them to cancel each other out).

Note that the original paper mentions, but neither explains, nor correlates a shift observed in the $D_3$-$D_4$ detector's collapsed waveforms' peaks (and detector $D_4$ is not even featured on the schematics in the original paper, as seen in Fig. 1). The $D_3$-$D_4$ peak shift may be much more important, than it seems. It still indicates that some of the observed key phenomenon (i.e the apparent 180°(or $\pi$) phase shift for the $D_1$-$D_2$ detectors) is a consequence of the original topology's 'eraser-paths' redundant symmetry, independent of *principle*. Furthermore, if shifted, the $D_3$-$D_4$ joint detections distribution curve must feature two statistical maximums, which indeed, could make it partially distinguishable from $D_1$-$D_2$ joint detections, thus carrying an estimated $\sim 10^{-1}$ – $10^{-3}$ (non-zero) bit/signal detection information on the idler's later choice, in *advance* of the idler's registration, and *without* the coincidence counter.

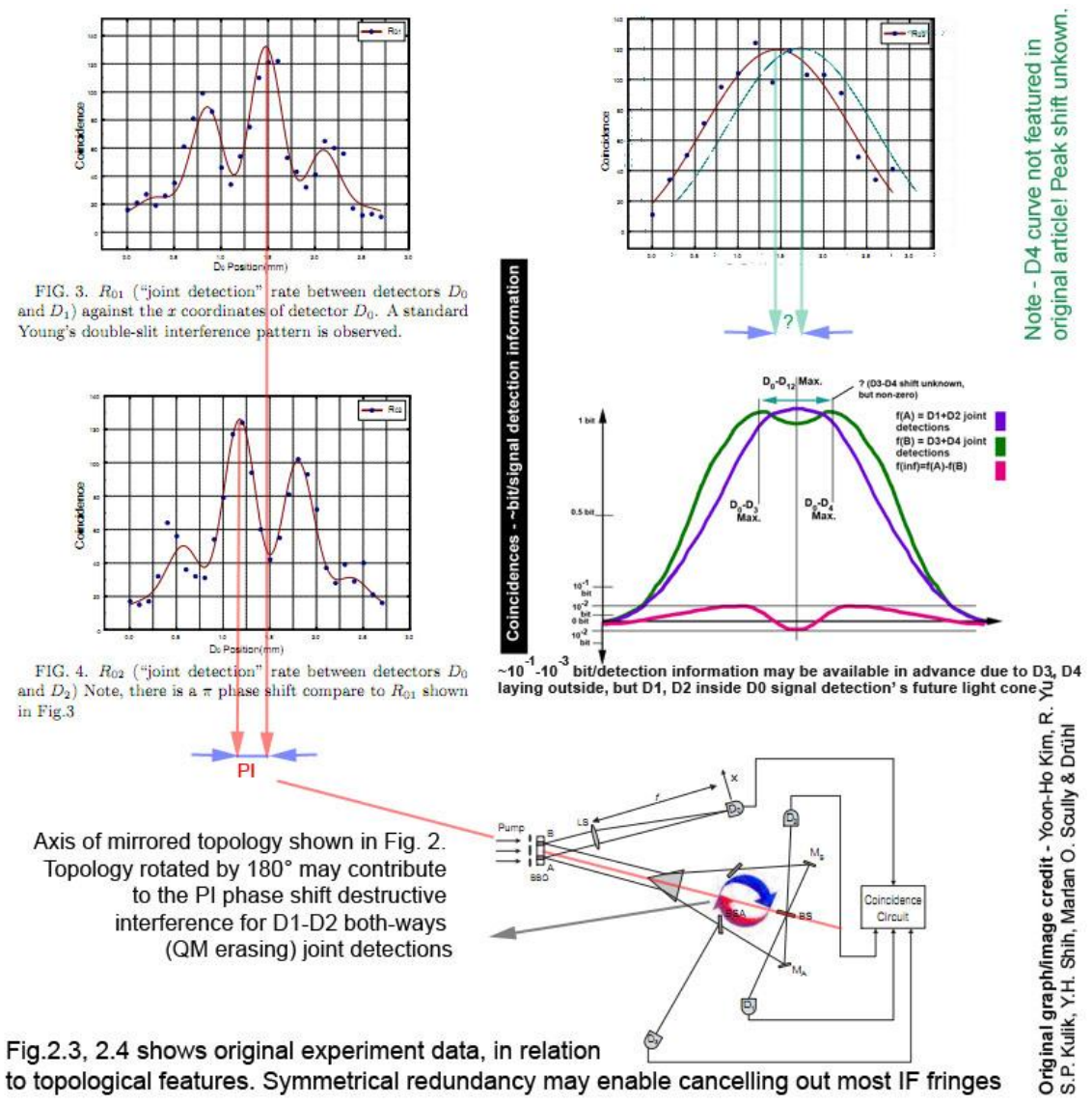

Fig. 2. illustrates the questionable shifts. Partial advance-in-time information (shown in right middle graph) may be available because $D_3$-$D_4$ seems to lay outside, while $D_1$-$D_2$ inside $D_0$'s future light cone.

However, if original topology is to scale, There may also be a much stronger principal – QM may not allow paradoxes. And to avoid possible violation of causality, it must hide information on the future from the local observer, ensuring that it can not intervene to change the already observed future.

## Explaining topological loopbacks

If we carefully analyze the original setup of the DCQE experiment by Yoon-Ho Kim, R. and by Marlan O. Scully, we must realize that the idler photon's detection (*at least* $D_1$ and $D_2$) lays *inside* the future light cone of the signal's ($D_0$) detection (and its local observer's, if any).

This is all because mirrors are used to alter the course of the idlers, and not letting them propagate along the light cone's edge. They are redirected close enough to $D_0$.

Therefore, if the local observer at $D_0$ gains knowledge on the idler twin's future choice, he (or she) can still (both theoretically, and also technically) intervene to intentionally change the experiment's setup, and by this violate causality and realize a paradox (by changing already observed future).

[3] Note that the coincidence counter is only needed for the local observer to prove the existence of the correlation. If it was possible, for the local observer, to interpret the signal photon's detection in $D_0$, in relation to its idler's later choice, quantum-level information on the future could be obtained. Of course, it is impossible in the original setup; we now examine the constrains.

If causality stands, QM should not let this to happen, so it must find a way to hide the already existing information on the twin's delayed choice from the local observer at $D_0$.

In the original setup, as we had already shown, this manifests itself in the 180° (complementary) phase shift of the $D_1$-$D_2$ eraser detectors[4].

**Avoiding the paradox**

If the observer can get knowledge of the future, but can not, even in principle do anything to change it, causality is *not* violated.

One way of ensuring this would be to place the target of observation on the hypersurface of, or outside the future light cone (relative to the $D_0$ detection and its observer).

The easiest way to achieve this is to let the idler twins propagate in a straight line, - preferably in outer space - without any mirrors or beamsplitters altering their path, or course[5] (before detection).

A photon, propagating freely in space, by the speed of light (*c*) will always be found on (or fluctuating around)[6] the hypersurface of the light cone. Therefore, if will interact with anything that causes erasing (or marking) which-way path information, and the local observer gains knowledge of that by observing the local ($D_0$) detection, he (or she) can do nothing – even in principle – to change it. The interaction's space-like distance would be exactly as far away as achievable by light; this way causality could not be violated, and no paradoxes should occur.

Therefore, gaining information from that special hypersurface of the spacetime should be possible for the local observer.

It should be emphasized, that the local observer (at $D_0$) would *not* need to wait (i.e. 2 million years) for the idlers to reach a distant target (i.e. in Andromeda Galaxy). Information on the idler's future fate could be *immediately* available by local signal ($D_0$) detection.

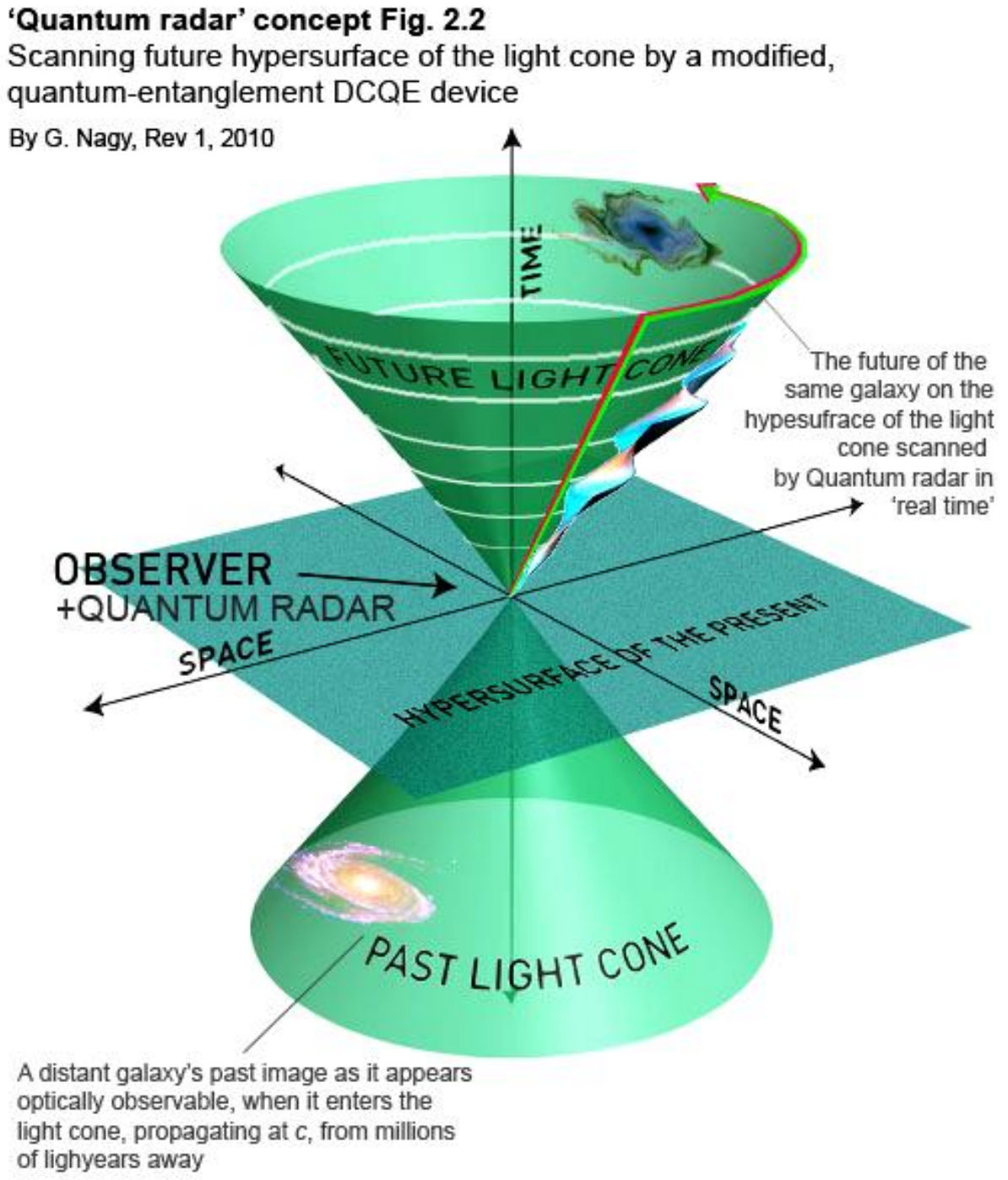

Fig. 3-2.2. illustrates concept of such remote sensing.

**Testing the theory**

If we are to exploit this feature, we propose to simply remove all mirrors, beamsplitters and coincidence counters from the 'idler' part of the experiment.

Then, for the 1st test, copy the 'signal' setup symmetrically to where the idlers part was. Set up the $D_1$ detector exactly as $D_0$.

We predict that the outcome should be an interference pattern on both screens (whether or not a 180°or π phase shift occurs, although likely, is now obviously irrelevant).

Now, for step 2, take the 'idler' part and the $D_1$ screen very far away from the $D_0$ screen and the local observer. We predict, that – even though interference patterns on both screens may be rotating symmetrically – the type of the patterns should not change.

For step 3, we introduce a remote triggering mechanism at the distant ($D_1$) screen that can change the setup very fast (ie. by opto-electronics) to detect or erase the which-way path. The remote triggering mechanism, would be activated by a normal (ie. radio) signal that travels by *c*.

Fig. 3 shows experimental setup schematics needed for testing all three (future, present and past) hypersurface quantum signaling.

Note that focusing probability waves, with adaptive optics / mirrors to scan larger distances would be technically very challenging, but theoretically possible (even for cosmological distances).

---

[4] If it should turn out that any modifications in the topology, without placing the the distant ($D_1$..$D_n$) detectors outside the local observer's light cone, possible, it would give way for a violation of causality, ie. the future retrocausally changing the past. What we are trying to show is that it can be avoided.

[5] If we need to introduce mirrors in the idler's part, we can still place the $D_0$ detector, along with its local observer, outside the light cone by introducing one more mirror for the signals which reflect the photons to the opposite direction (of the idler's propagation). This way, we should still be able to obtain information without violating causality or invoking paradoxes.

[6] Please see the 'Multi-dimensional hyperwaves theory'

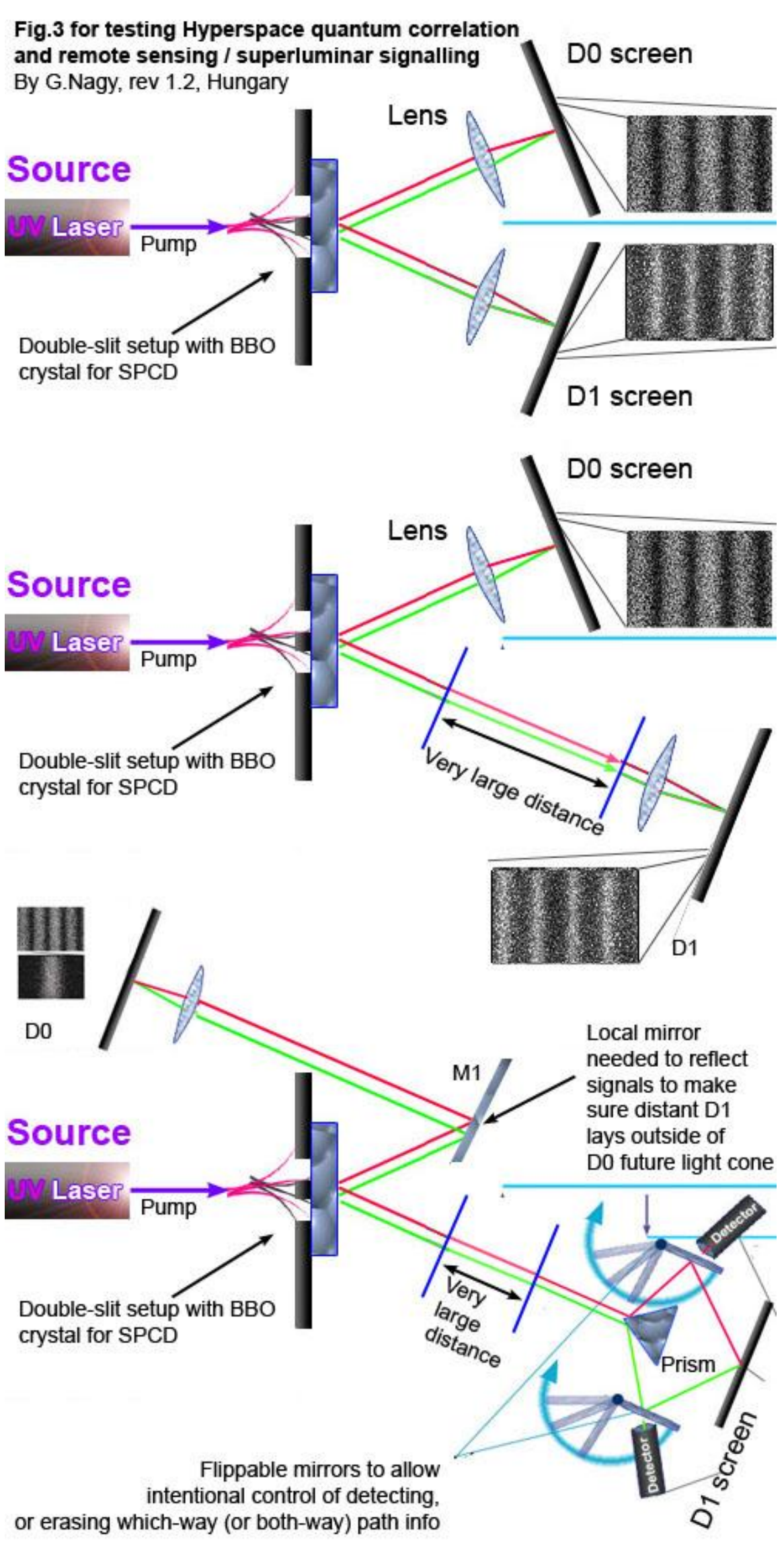

If our theory is correct, the local wavefunction should start to collapse immediately when we send out the signal from $D_0$ – that is, without having to wait for the triggering signal to reach its distant target (the $D_1$ screen). Why? Because the signal photons in $D_0$ are entangled with the future of their twins (in our frame of reference), and the which-path marking will start exactly at that point in the future when the triggering signal reaches the distant setup.

While on the other side, the idlers photons at $D_1$ are entangled with the past of their signals (in our frame of reference).

If we send another signal, to restore the original setup (erase which-path information), the interference fringes should start to reappear immediately in $D_0$.

From all this – if it works – we can conclude that an observer, who choses to gain knowledge of the which-way path, can only see the past (it would be consistent with our optical observations of the universe). An observer who chooses to erase the which-way path (thus preserving the wavefunction) can only see the future (in his or her local frame of reference).

But since particles are entangled, if any side detects the which-way path, the collapse also happens on both ends. If they both measure, there's no information available for any of them, elegantly avoiding another paradox.

If the experiment does not comply with predictions, we must assume that either the signal and/or idler photons are, indeed, *not* all propagating by the speed of light; and the Theory of Relativity may be challenged[7].

For such case, we propose two other solutions, both of them implicating that gaining information from outside the light cone (hyperspace) may still be possible.

## Interacting with the 'present' hypersuface

According to Heisenberg's Uncertainty Principle[4] , an infinitely accurate marking of position implies infinite uncertainty in momentum.

Photons do not have mass, so we could only apply the Principle to speed. Detecting a photon's exact position would lead very high uncertainty in its speed (ie. could reach many times of *c*), possibly infinite speed (in case of infinitely accurate position detection). Of course, in case of a 'normal' photon this is meaningless to discuss, since accurate detection of a photon's position can only be carried out by destroying the photon at the same time.

But with entangled photons, something very different may happen. It may be possible that the both the signal's, and the idler photon's speed will be 'infinite', or very high (many times that of *c*), if the position of at least one of them will be very accurately measured before detection of both of them (in the local observer's frame of reference).

---

[7] The original paper by by Yoon-Ho Kim, R. Yu, S.P. Kulik, Y.H. Shih, designed by Marlan O. Scully & Drühl fails to provide actual details on the measured time difference of the signal detection(s) $D_0$ detector, and the detection(s) of the idler(s) in the $D_1$..$D_4$ detectors. The paper seems to presume that both signal and idler photons will definitely propagate by the speed of light (*c*), so it only introduces a simple calculation, stating that there should be a constant, 8 *n*Sec delay between $D_0$ and $D_n$ detectors, as they are approx. 2.5 meters apart from $D_0$ (optical path). Note that in itself it can clearly not be true, even if photons indeed are propagating by c, since the $D_0$-$D_3$ and $D_0$-$D_4$ optical paths are significantly shorter than $D_0$-$D_1$ and $D_0$-$D_2$ paths. Missing data may be the deciding factor.

In this case, the wavefunction of the signal photons would not be dependent on the future of the idlers. It would, instead, be dependent on the *present* (or very close to the present) hypersurface interaction of the idlers in spacetime.

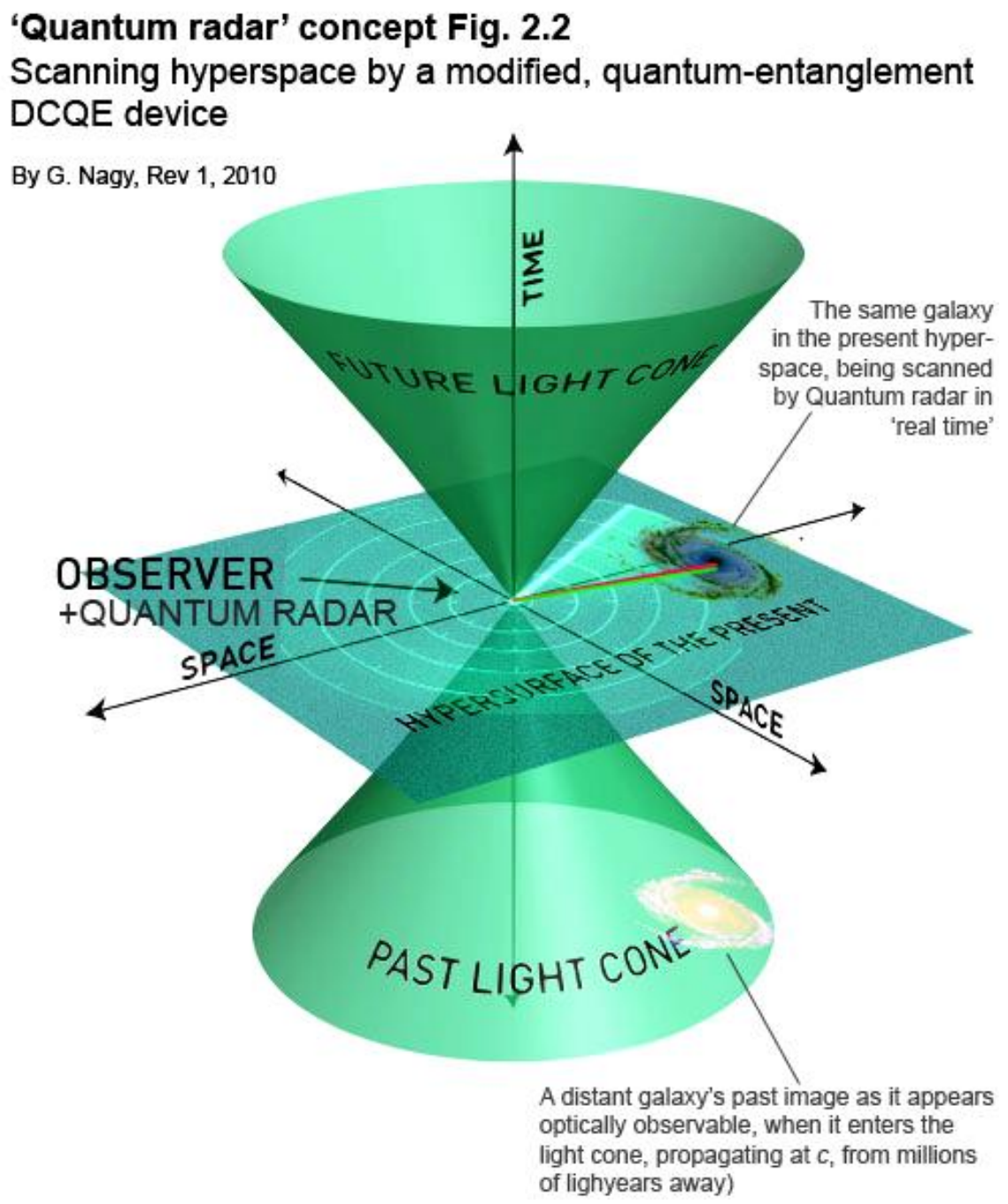

Figure 4-2.2. illustrates concept of such sensing.

Testing the theory would be easy.

With the same, remote setup of the $D_1$ screen and with an intentionally controllable marking or erasing of the which-way path, the local observer in $D_0$ could send out a normal triggering signal to start marking which-way path.

Local interference pattern should start to be collapsing when the normal triggering signal reaches its destination (travelling by *c*).

In this case, we would be interacting with the entangled particle in present hyperspace.

Yet, there is one more alternative to discuss.

## Past-hyperspace interaction

Uncertainty in the speed of light need not necessarily result in speeds higher than c for both particles. There is one more way of ensuring that detection happens at the same time even in he observer's frame of reference.

For this, the speed of the 'signal' particle may be forced to be lower than c; while the idler particle would need to move faster than *c*.

The exact ($v_s$, $v_i$) speeds would be easily expressable by the ratio, or difference of the optical length of paths (between the SPDC source and the $D_0$, $D_1$ screens, respectively).

In this case, the wavefunction of the D0 photon would be dependent on the past-hyperspace interaction of the idlers, where the hypersurface's angle (between the past light cone and the present) would also be defined by the ratio of the optical paths of $v_s$, $v_i$.

## Possible practical uses

Each of the theories above offer the obvious ability to realize superluminar communication, as well as remote sensing (mapping) quantum properties of unknown regions of spacetime[8].

Fig. 5 shows the concept of detecting a Solar burst in advance

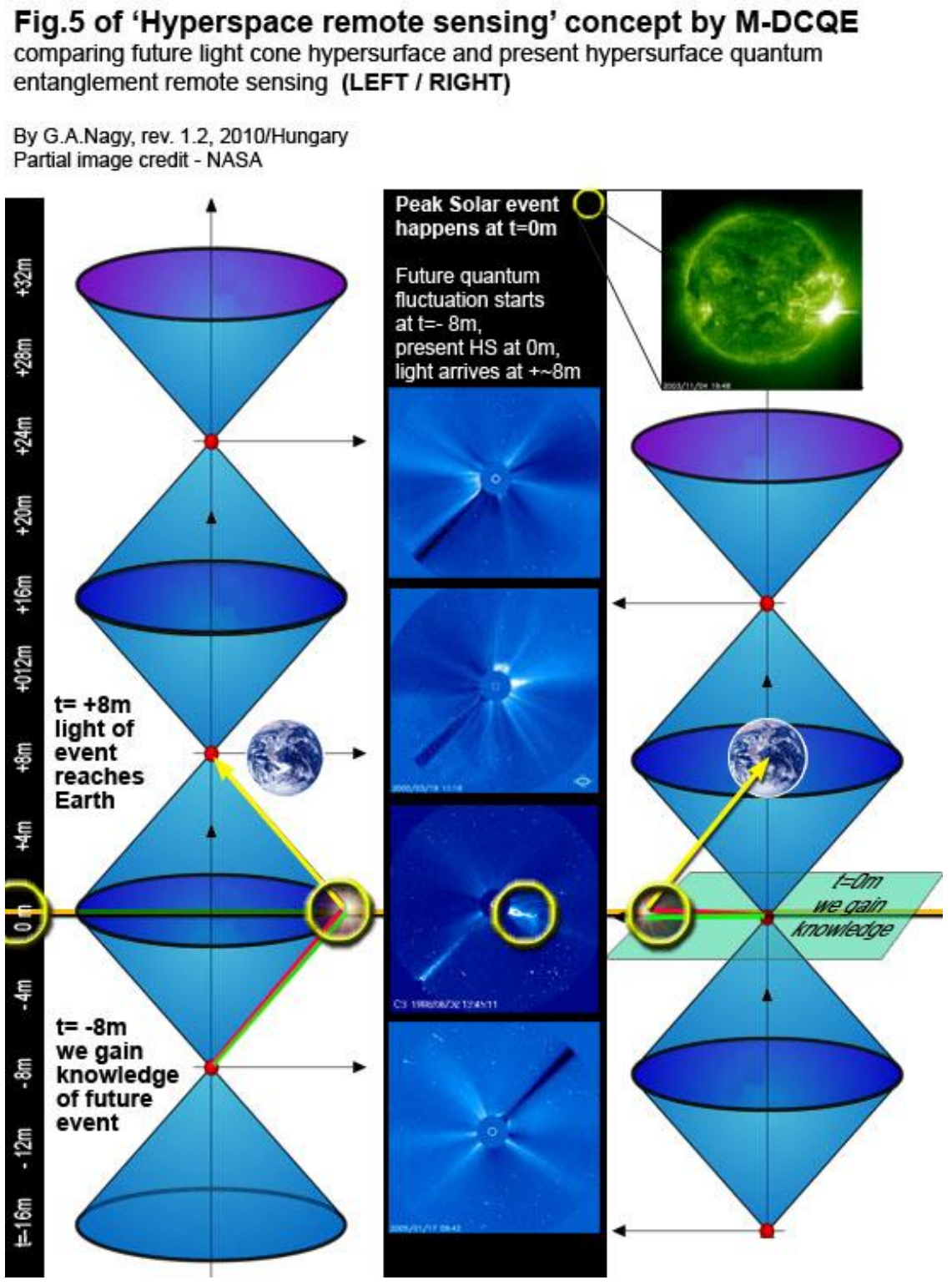

Also note that such 'remote sensing' could reveal information on cosmological events which have not yet entered the normally observable part of the universe, of the light cone (i.e. are happening 'real-time', simultaneously with the distant observation.)

---

[8] When using the modified DCQE for measuring remote quantum properties of spacetime, the pattern in the $D_0$ screen will be dependent on whether the interaction (of unknown depth, or distance), is such that it 'erases' or locally exploits ('marks') which-way path. When scanning natural or artificial objects – such as gas, liquids, metal, rocks or plasma, one could not hope to receive either a totally intact interference pattern (fringes), nor a completely collapsed one. The local observer would be likely be receiving very fine fluctuations of the pattern, somewhere between half-collapsed and half-intact fringes.

We could, for example, remote-sense a solar flare burst of the Sun, in real time (or even in advance), not having to wait for light of the event to reach us.

Fig. 6. shows a superluminar signaling setup

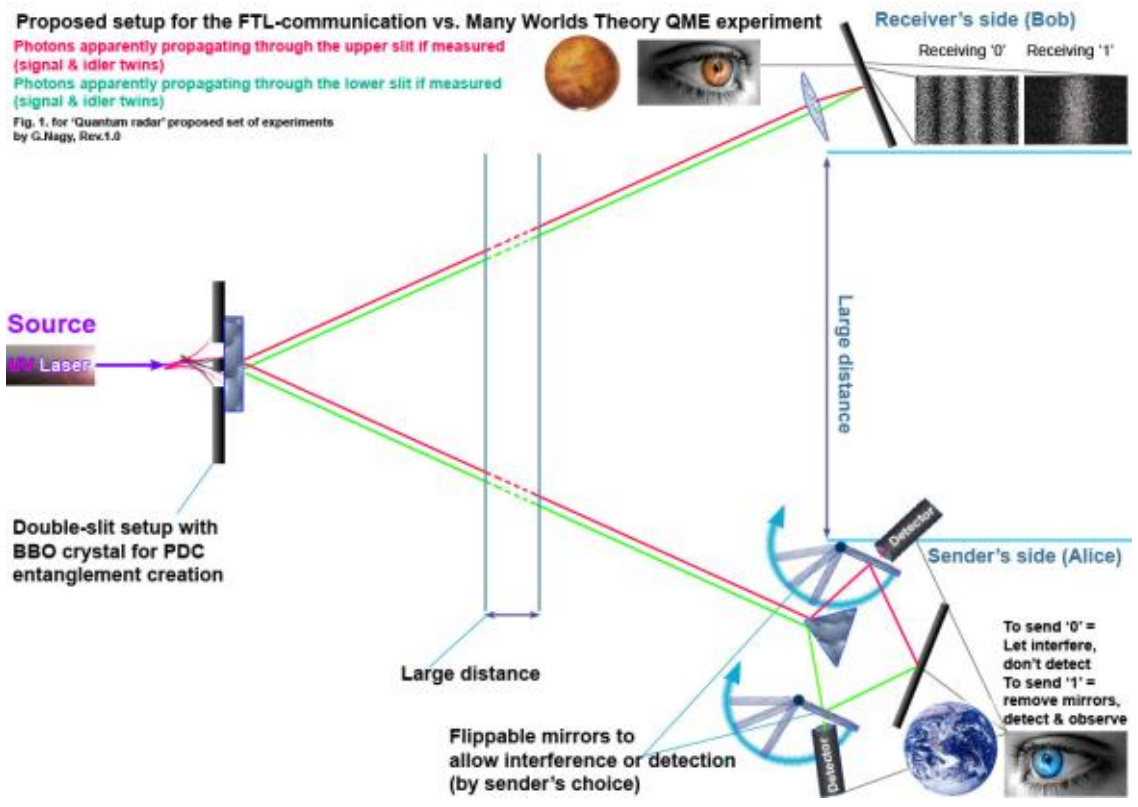

Please note that the above realization is a special, symmetrical subcase of the DCQE which does not even require entanglement over the dimension of at time at all. It also does not violate Relativity since if information travels to its past on one side, it travels to its future on the other. Thus, it arrives 'real-time' present for the distant receiver.

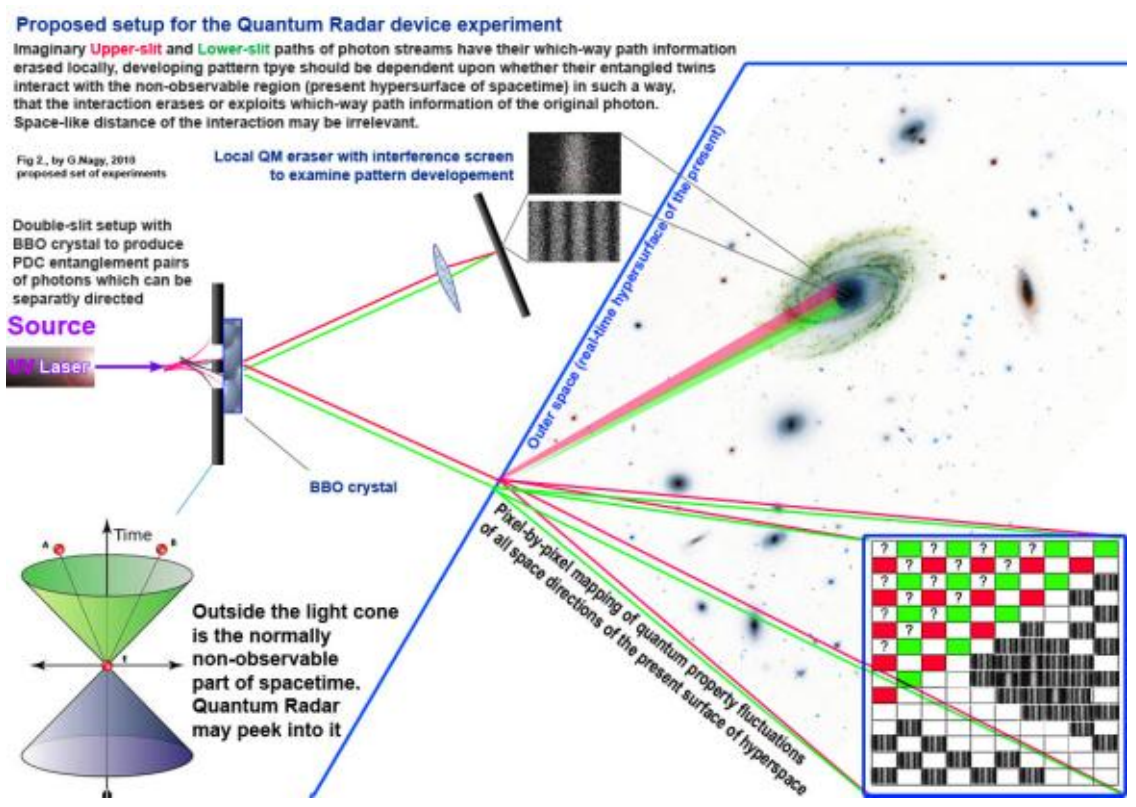

Fig. 7. shows a visualized concept of such mapping of hyperspace

Superluminar communication and remote sensing would also be very useful if we enter the interplanetary or interstellar area, where normal communication could take minutes, days or even years.

## Conclusion

We theorized that obtaining quantum-level information from either the past, present or future hypersurfaces does not necessarily violate causality, and therefore should be considered possible.

We based our conclusion on the results of a classic DCQE experiment, where we had shown that reason for not being able to extract meaningful information before detecting both the idlers and signal photons may simply be a failure of the local loopback topology, with the detection screens located in the light cones of each other, capable of violating causality and causing a paradox. Also, the symmetrical mirroring D1-D2 detectors, leaving the chance to cancel interference fringes out, can simply be avoided.

Changing the topology and removing or counter all optical loopbacks should also remove such limitations in principle.

Testing the theory is possible with today's technology already available in well equipped quantum-optical laboratories; yet if any chance of success or experimental implication shows predictions could be correct, real use of such remote-sensing equipment would be in space.

For humanly observable results, a distance of at least 0.1-1 lightseconds between the local observer (sender), and/or the scanned objects (or receivers) would be desirable.

Quantum property map of hyperspace could be scanned just like background microwave radiation; showing the optically non-observable regions of our universe.

*Note*

**Appendix A** contains a short introduction of the Multi-dimensional Hyperwaves theory, and its implications in relation of hyperspace remote sensing devices theorized in our article.

Contact information on he author(s) is also available in App. A.

# Appendix A

## The 'Multi-dimensional Hyperwaves' Theory[9]

With its implications on possible hyperspace, or future hypersurface remote-sensing devices

*By Gergely A. Nagy, 2010, Hungary*[10]

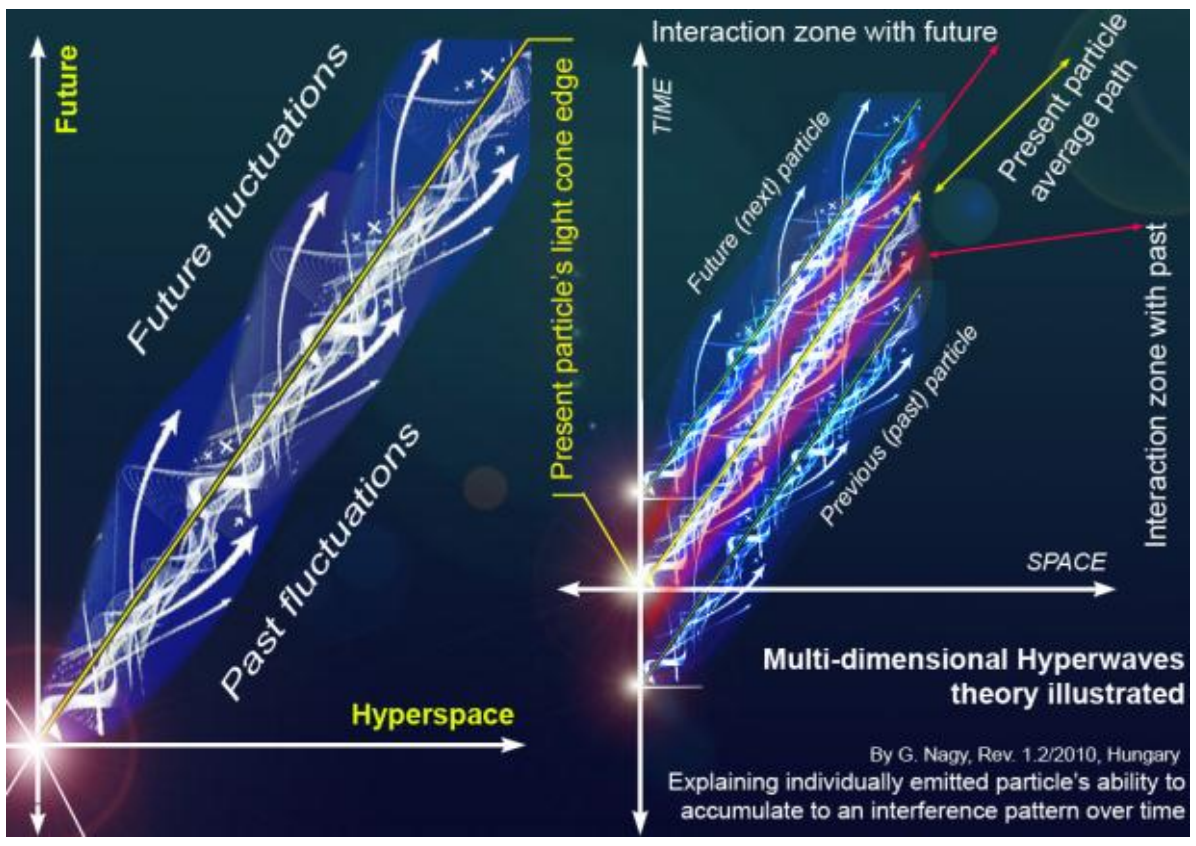

We theorize that the probability wave of the individual particles (emitted from the source, but not yet detected, i.e. in a double-slit setup) not only oscillates almost freely in 3-dimensional space, but also in the dimension of Time. Therefore, the individual particles can easily interfere with the next, and the previous particles in the repeated process of emissions as well, interact with each other (in future, and past hyperspace), and return to create the interference pattern in the present.

This means, that even though an individual particle has already been detected, its probability wave still exists in its relative future (in the 'present'), and the next emitted particle can interact with it (as its probability wave also fluctuates into its relative past (in hyperspace)).

We propose that this theory (which we call *'Multi-dimensional hyperwaves'*, referring to the individual particle's freedom being extended to Time, higher dimensions and maybe even to hyperspace) is much simpler, yet provides a more elegant way of explaining the interference development phenomenon than, for example, the elementary waves theory.

And this theory, however extraordinary and controversial it may sound, should not create any paradoxes, after all (even if it may seem to imply an already-existing future, but it does not.)

Particles can even interfere with both next & previous instances of individual emissions, and if we are to stop the experiment at will (no future individual emissions), wavefunction should still be preserved, at least partially by past-hyperspace interactions with already emitted, individual particles in the sequence.

Our proposal may be examined experimentally by, for example, carefully increasing and decreasing the time between each individual emission, and looking for statistical anomalies (or simply, some type or kind of changes in the distribution of particle manifestations) in the evolution of the interference pattern.

If such correlation is revealed, we propose, a *not-yet named* constant could be derived that would describe the functional dependence (or simply linear ratio) between the units we use to measure space, and time, as we know it.

### Implications on Hyperspace remote-sensing

Fluctuation of the probability wave not only in space, but also in the dimension of time means that it is only the statistical (mean) average of the idler photons apparent path, that is lying on the given hypersurface. And since idler photons not always staying on the given hypersurface, their first interaction on remote spacetime may happen outside of it. So some of collected data from hyperspace (or hypersurface of the future light cone) may indeed also originate slightly off course.

If the theory is correct, the most crucial implications would be considering the possibility that – when scanning along the hypersurface of the future light cone – we may obtain information from *within* the light cone [relative to the local observer]. This would, unless countered, threaten a violation of causality.

However, we theorize that fluctuations into the opposite direction in time with uniform distribution will ultimately cancel out, and extractable information will always reflect average quantum properties alongside the statistical average (or mean) path, defined by the probability wave of the idler photon's apparent in-line propagation.

---

[9] (Original 'MDHW' theory presented in a paper by Idokep.hu., Ltd. science columns, article id. 984.)

[10] Experimenters interested with Quantum-Optics lab access are welcome to contact author(s) by gergely@idokep.hu, idokep@idokep.hu, +36/70-9426259, +36/20/448-2180, Idokep Kft., Bartok Bela str. 65/b., 1224 Hungary, to test theories above for possible joint publication and R&D. Any such contact is much appreciated.